\documentclass[sigconf]{acmart}
\usepackage{graphicx}
\usepackage{hyperref}
\usepackage{natbib}
\usepackage{graphicx}
\usepackage{hyperref}
\usepackage{enumitem}
\usepackage{lineno} 
\usepackage{makecell} 
\usepackage[table]{xcolor}
\usepackage{subcaption}
\newcommand{\refsec}[1]{Section~\ref{#1}}
\newcommand{\reftab}[1]{Table~\ref{#1}}
\newcommand{\refequ}[1]{Equation~\ref{#1}}
\newcommand{\reffig}[1]{Figure~\ref{#1}}

\newcolumntype{C}[1]{>{\centering\arraybackslash}m{#1}}

\renewcommand\footnotetextcopyrightpermission[1]{}

\begin{document}

\title{Synthetic Wastewater Epidemiology Data Generation using Patterns-of-Life Simulation}
\author{Hossein Amiri}
\orcid{0000-0003-0926-7679}
\email{hossein.amiri@emory.edu}
\affiliation{%
    \institution{Emory University}
    \city{Atlanta}
    \country{USA}
}

\author{Mohammad Hashemi}
\orcid{0009-0005-1608-7213}
\email{mohammad.hashemi@emory.edu}
\affiliation{%
    \institution{Emory University}
    \city{Atlanta}
    \country{USA}
}

\author{Akshay Deverakonda}
\orcid{0009-0008-9899-271X}
\email{Akshay.deverakonda@emory.edu}
\affiliation{%
    \institution{Emory University}
    \city{Atlanta}
    \country{USA}
}

\author{Joon-Seok Kim}
\orcid{0000-0001-9963-6698}
\email{joonseok.kim@emory.edu}
\affiliation{%
    \institution{Emory University}
    \city{Atlanta}
    \country{USA}
}

\author{Yuke Wang}
\orcid{0000-0002-9615-7859}
\email{yuke.wang@emory.edu}
\affiliation{%
    \institution{Emory University}
    \city{Atlanta}
    \country{USA}
}

\author{Andreas Z{\"u}fle}
\orcid{0000-0001-7001-4123}
\email{azufle@emory.edu}
\affiliation{%
    \institution{Emory University}
    \city{Atlanta}
    \country{USA}
}

\renewcommand{\shortauthors}{Amiri, et al.}
\begin{abstract}
Wastewater contains rich biological signals that can be used to monitor population health, track infectious diseases, and detect emerging outbreaks. Pathogens and other biomarkers in sewage provide a unique, noninvasive view of disease prevalence at the community level. However, extracting these signals requires extensive field sampling, laboratory analysis, and expert interpretation. Consequently, wastewater-based epidemiology (WBE) datasets are scarce, geographically fragmented, and rarely released as open data. Even when available, existing datasets typically cover short time periods and limited geographic regions, restricting their usefulness for method development, benchmarking, and large-scale modeling.
To address this gap, we present an application of an existing patterns-of-life simulation framework for generating synthetic infectious disease and wastewater pathogen datasets for wastewater-based epidemiology. Our approach extends a patterns-of-life simulation framework by using it as a model of human mobility and behavior while incorporating disease transmission and pathogen shedding dynamics. The resulting framework generates high-resolution spatial and temporal datasets capturing infection dynamics, mobility behavior, and wastewater-associated pathogen signals.
We release a fully simulated dataset containing check-in records, social network links, infection states, pathogen loads, and ground-truth disease transmission information. These data support controlled experimentation for outbreak detection, source localization, resource allocation, surveillance strategy design, mobility-aware wastewater analysis, and targeted public health interventions.
We provide datasets for Fulton County and demonstrate that the framework generalizes to other regions by generating data for any city with available OpenStreetMap information.
By making large-scale synthetic wastewater epidemiology datasets readily available, this work lowers the barrier to research in outbreak monitoring, epidemiological modeling, and wastewater surveillance decision support.   
\end{abstract}

\maketitle
\pagestyle{plain}
\section{Introduction}
\label{sec:introduction}
Public health decision making heavily relies on population-level surveillance observations, such as number of cases reported or pathogen concentration in wastewater samples. Understanding how human behavior, mobility, and physiological processes impact surveillance observations is a critical problem at the intersection between spatial data science and public health. These factors govern how individuals move through space, where they spend their time, and how they act and interact, which aggregate into observable signals from contact events that drive disease transmission to numbers of pathogens circulated in wastewater.~\cite{amiri2026we,delapaz2025integrating}. Yet real-world datasets linking behavior and mobility to surveillance observations are noisy, incomplete, and constrained by privacy, scale, and sampling bias, which limits their suitability for controlled experimentation and systematic evaluation~\cite{medema2020implementation,wade2022understanding,kong2025simulated}. 
\begin{figure*}[t]
    \centering
    \includegraphics[width=0.8\linewidth]{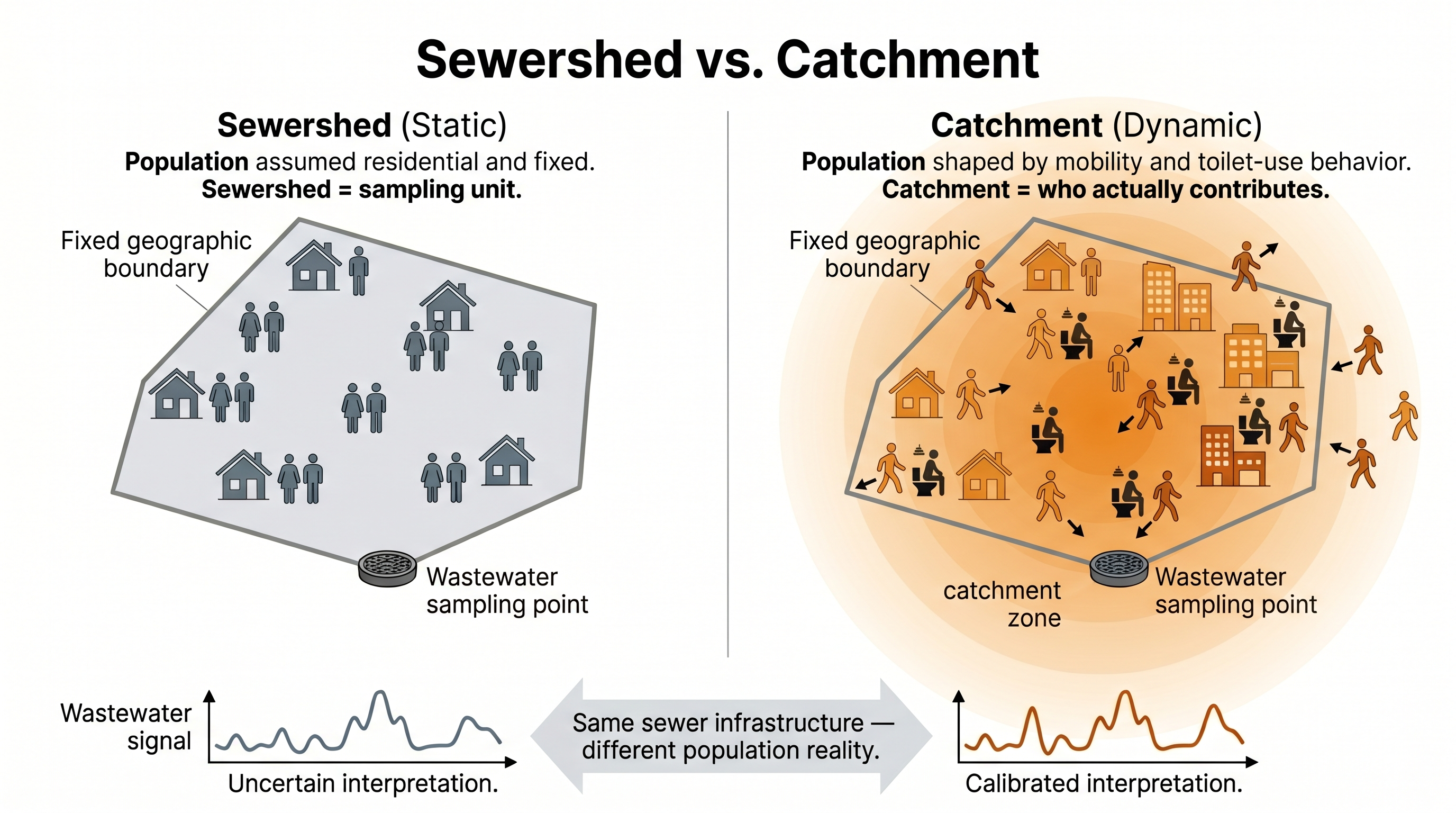}\vspace{-0.2cm}
    \caption{The sewershed represents a fixed geographic sampling unit with an assumed residential population, whereas the catchment represents the dynamic population that actually contributes wastewater to the sampling point. Accounting for mobility and toilet-use behavior can improve interpretation and calibration of wastewater signals.\vspace{-0.05in}}
    \label{fig:sewershed-versus-catchment}

\end{figure*}
Recent advances in patterns-of-life simulation demonstrate that behaviorally grounded agent-based systems can generate large-scale, reproducible datasets of human movement, behavior, and interaction that preserve temporal rhythms, spatial structure, and individual heterogeneity~\cite{amiri2024patterns,amiri2024geolife+}. These systems produce synthetic trajectories that encode routine activities such as commuting, dining, and social visits, yielding rich spatiotemporal records of urban life~\cite{amiri2024patterns}. Such datasets enable rigorous evaluation of learning, detection, and inference frameworks and support downstream tasks including large-scale mobility analysis, anomaly detection, and the study of social and semantic trajectory structure~\cite{zhang2024transferable,zhang2024large}. 
Crucially, this spatial explicitness---knowing where and when each behavior occurs---is what makes such data suitable for public health inference: contacts arise where people gather and pathogens are shed wherever people use toilets, so the same population can generate very different surveillance signals.

Since the recent COVID pandemic, wastewater-based epidemiology (WBE) has emerged as a powerful population-level tool for disease trend inference and early warning for disease outbreaks~\cite{thompson2020making,mao2020potential,wang2022early,medema2020implementation}. Yet its interpretability is constrained by a simplifying assumption the field has long acknowledged: most WBE studies treat the contributing population as residentially fixed within a static sewershed (the physical drainage area of a sampling point), even though the actual catchment (the population whose excreta reaches a sampling point during a given time period) is dynamic. The catchment diverges from the residential population within the sewershed whenever people move and use toilets away from where they live. Such a divergence is continually reshaped by mobility, social mixing, and toilet use behavior. This assumption that catchment is approximately equal to the residential population within the sewershed holds at coarse spatial scales and stable periods but breaks down precisely where surveillance is most consequential: at upstream building- and neighborhood-level resolution and during periods of unusual behavior such as holidays, mass gatherings, or the early days of an outbreak. Recent agent-based frameworks have begun to integrate disease dynamics, mobility, and wastewater processes~\cite{delapaz2025integrating,schmid2025integrative}, but none provides datasets that jointly capture mobility behavior, disease progression, and spatially resolved wastewater signals within a unified geospatial environment. Our work addresses this gap by generating such datasets---coupling agent mobility, infectious disease dynamics, and toilet use events---to enable controlled study of how dynamic catchments shape the signals observed in wastewater. A comparison between the sewershed and catchment concepts is shown in~\reffig{fig:sewershed-versus-catchment}, highlighting the importance of human mobility in wastewater epidemiology.

In this paper, we build on the framework introduced in our prior work~\cite{amiri2026we}, extending patterns-of-life simulation. The Patterns of Life model~\cite{amiri2024patterns} is an agent-based system grounded in Maslow’s hierarchy of needs~\cite{maslow1943theory}, where each agent’s daily behavior emerges from the continuous satisfaction and accumulation of physiological and social requirements. We introduce defecation as an explicit physiological need and couple it with infectious disease dynamics, so that agents defecate, in a behaviorally consistent manner, at the toilet locations where they spend time---enabling the simulation to generate wastewater signals and support WBE. With this foundation, our main contributions are as follows:

\begin{itemize}
    \item We utilize the extended Patterns of Life simulation framework to incorporate defecation as an explicit physiological need that is coupled with Susceptible-Exposed-Infectious-Recovered (SEIR) infectious disease dynamics to simulate hypothetical disease outbreaks and associated toilet use behavior and pathogen shedding.
    \item We present a synthetic geospatial data generation pipeline that integrates human mobility, health states, and behavior-driven physiological signals, enabling controlled experimentation across mobility analysis, epidemiology, and wastewater-based surveillance.
    \item We provide 84 datasets generated across diverse scenarios varying in disease characteristics, population behaviors, and intervention strategies, supporting evaluation of public health inference and surveillance systems.
    \item We present analysis of the generated datasets, demonstrating their utility for benchmarking models in mobility analysis, epidemiology, 
\end{itemize}

We clarify that our framework does not simulate the physical transport of pathogen loads through the sewer network, such as travel time, decay, or sampling point placement. Instead, our work focuses on modeling human behavior and mobility, including defecation behavior based on Maslow’s hierarchy of needs, where pathogen shedding at toilets is explicitly captured. Modeling pathogen fate in the physical sewer network is beyond the scope of this study and can be addressed in future work or as a separate data-processing pipeline.

The remainder of the paper is organized as follows.
In~\refsec{sec:related}, we review related work on patterns-of-life simulation and wastewater-based epidemiology.
~\refsec{sec:methodology} presents our extended simulation framework and data generation pipeline.
In~\refsec{sec:results}, we describe the experimental results including generated datasets, scenarios used for data generation, configuration details for each dataset, and visualizations of the results.
~\refsec{sec:reproducibility} discusses reproducibility, and data availability.
Finally,~\refsec{sec:conclusion} concludes the paper and outlines directions for future research.

\section{Background and Related Work}
\label{sec:related}
In this section, we provide background information and review previous work on patterns-of-life simulation and wastewater-based epidemiology.
\vspace{-0.1in}
\subsection{Patterns-of-Life Simulation}

The Patterns of Life simulation is an established agent-based modeling framework that represents realistic everyday human activity within real or synthetically generated urban environments by simulating how individuals live, move, work, and interact over time. Agent decision-making is theory-driven and grounded in Maslow’s hierarchy of needs~\cite{maslow1943theory}, with agents balancing physiological, safety, and social requirements such as sleeping, eating, working, financial stability, and social interaction to maintain well-being. Time advances in discrete steps mapped to real clock time, enabling the model to capture daily and weekly rhythms, weekends, and calendar events, while agents reside in homes, commute to workplaces or schools, dine at restaurants, and socialize at recreational venues~\cite{kim2020location}. The simulated environment is initialized using real geographic data, typically derived from OpenStreetMap, including buildings, building units, and walkway networks that support realistic routing and mobility. Cities are divided into neighborhoods with buildings assigned functional types, capacities, costs, and attractiveness values sampled from specified distributions to reflect plausible urban compositions. The population consists of individuals and families with heterogeneous demographic, economic, and behavioral attributes such as age, education, income, interests, and social needs. Social and economic dynamics are explicitly modeled through evolving family, friendship, and workplace networks, where interactions strengthen ties and inactivity causes decay, alongside detailed representations of employment, income, expenses, and household finances~\cite{amiri2024patterns}. The design, validation, and extensions of this framework have been documented in prior studies~\cite{ kohn2023epipol,kong2025simulated}.
The framework has been extended into HD-GEN, an integrated high-performance data generation system~\cite{amiri2026hd} that provides pipelines for calibration against real-world datasets, structured outputs, and large-scale generation of synthetic mobility data, using the Patterns of Life simulator as its core component. HD-GEN can generate data for any geographic region. It queries geographic information from the Overpass API~\cite{overpass} to construct a map for the simulation. It also includes built-in calibration based on a genetic algorithm to align simulation outputs with empirical mobility characteristics such as trip counts and spatial dispersion~\cite{amiri2024geolife+}.

\vspace{-0.1in}
\subsection{Wastewater-Based Epidemiology}

Wastewater surveillance (WWS) is a public health monitoring approach that analyzes sewage to detect and track diseases and other health indicators at the community level, offering a population-wide snapshot of health that complements existing individual clinical testing; its foundations date to mid-20th-century efforts that successfully monitored poliovirus circulation through sewage, demonstrating the feasibility of tracking infectious diseases via wastewater~\cite{paul1941virus, trask1942periodic, rhodes1950poliomyelitis}. In practice, wastewater is collected from strategic points in sewer networks, commonly treatment plants or targeted upstream sites, and analyzed for genetic material or biomarkers, enabling inference of disease presence  across the served population. WWS provides population-level coverage without individuals having to seek health care,~\cite{thompson2020making}, captures both symptomatic and asymptomatic infections~\cite{medema2020implementation}, often functions as an early warning system for outbreaks~\cite{mao2020potential, wang2022early}, is generally more cost-effective than mass testing, and reaches under-served groups with limited access to care, making it a powerful and equitable public health tool.

Wastewater-based epidemiology (WBE) extends WWS into a comprehensive analytical framework that uses wastewater measurements to infer population health, behaviors, and exposures, treating sewage as a pooled biological sample that reflects a community’s collective biochemical fingerprint~\cite{sims2020future, polo2020making}. Its impact became evident during the COVID-19 pandemic, when widespread deployment for SARS-CoV-2 enabled tracking of trends, early detection of surges, and hotspot identification at scales from cities to campuses and residences~\cite{scott2021targeted, wang2022early, wang2023case, hopkins2023citywide}. Beyond infectious diseases, WBE has been applied to quantify pharmaceutical use, estimate illicit drug consumption, assess nutrition via metabolic markers, and monitor environmental exposures~\cite{ort2010sampling, been2014population}. The approach offers objective, near-real-time, population data independent of healthcare access or testing behavior and reveals trends in hard-to-reach groups, including asymptomatic infections~\cite{medema2020implementation}; however, it yields only aggregated information, cannot identify individuals or precise locations~\cite{wang2023case}, and its linkage to true prevalence is affected by pathogen shedding variability, dilution, biomarker decay, toilet use, and population mobility~\cite{medema2020implementation, sims2020future, wade2022understanding}.
\vspace{-0.1in}

\subsection{Spatial Data and Applications }

Because each wastewater measurement aggregates the excreta of everyone defecates upstream of a sampling point, interpreting it is fundamentally a spatial problem: the signal reflects not only the spatial coverage of the underground sanitation infrastructure, but also the aboveground behavior of contributing population---where those people defecate and how they move in and out of a sewershed. Previous research has integrated spatial analytics and modeling into wastewater surveillance to study population movement and/or infection rates associated with observed signals.

In ~\cite{li2023spatio}  Bayesian spatio-temporal modeling were used to estimate wastewater concentration levels across England based on observed data from a set of wastewater treatment plants. Additionally, ~\cite{tang2022web} used a spatial decision support system to assist in building-level SARS-CoV-2 wastewater surveillance and identify spatiotemporal anomalies. A focus of the background literature has been using agent-based modeling to study spatial dynamics of wastewater systems and wastewater-based epidemiology. For example, ~\cite{delapaz2024modeling} constructed an agent-based model to simulate domestic wastewater variability at various spatial and temporal scales. ~\cite{0616456898e04215a978dfebbfd6f6f5} built on this work by using agent-based modeling to simulate both how populations move through wastewater catchment areas and how pathogen particles move through sewer networks. ~\cite{xiang2025wastewater} additionally used agent-based modeling to examine how wastewater surveillance can detect outbreaks depending on whether they start from sewered or non-sewered areas. Finally, ~\cite{bicker2025coupled} used agent-based modeling to reconcile outbreak and wastewater dynamics and examine spatial heterogeneity in disease prevalence at the city level.

One source of uncertainty is the day-to-day fluctuation in population sizes across different sewersheds ~\cite{boogaerts2024current}. To address it, population normalization methods rescale wastewater measurements for a more direct comparison across time and between sampling sites, although estimating catchment population size itself carries substantial uncertainty depending on the method used. Several such approaches include but are not limited to census data, mobile phone records, hydrochemical parameters, and various endogenous and exogenous biomarkers  ~\cite{boogaerts2024current}. 

These methods help to normalize populations to assist with comparison across sewersheds and enable comparisons of how populations change over time. However, there is still a gap in understanding spatial and temporal variation in populations that move through sewersheds. In particular, many studies of wastewater modeling assume defecation only at home, but few studies have examined wastewater signals under the alternative assumption that defecation also occur outside the home ~\cite{0616456898e04215a978dfebbfd6f6f5}. We address this gap by generating spatially explicit synthetic data that represents agent mobility, venue locations, disease status, and pathogen shedding events within a unified geospatial simulation framework. Each simulated agent moves through an environment built from real geographic structures, visits different venue types, interacts with other agents, and sheds pathogens at specific toilet locations it uses. Because every shedding event is spatially and temporally referenced, the resulting datasets support public health decision-making tasks that are difficult to evaluate with real data, such as hotspot detection, outbreak monitoring, source localization, surveillance site evaluation, and mobility-aware wastewater inference.  They can also drive an interactive visualization application that renders a simulated outbreak unfolding across a geographic area: agents moving between homes, workplaces, and venues; an infectious disease wave (e.g., SEIR dynamics) spreading through the population; and location-specific defecation-driven pathogen accumulating into the wastewater signal.

\section{Simulation Framework}
\label{sec:methodology}

We employ an agent-based geospatial simulation that builds on the Patterns of Life model which is validated and verified in~\cite{kim2020location} and the HD-GEN data generation pipeline~\cite{amiri2026hd}. The extended simulation framework, referred to in this work as \cite{amiri2026we}, integrates human mobility, social interactions, physiological events (e.g., defecation), and infectious disease transmission to generate comprehensive datasets. The HD-GEN framework \cite{amiri2026hd} supports large-scale scenario generation by configuring simulation parameters, calibrating behavior distributions, executing parallel runs, and processing outputs into structured formats. The augmented simulation incorporates modules for modeling this dataset is detailed in ~\cite{amiri2026we} and the source code of the original patterns of life and the extended wastewater-based epidemiology simulation
is available at \url{https://github.com/onspatial/wastewater-based-epidemiology-patterns-of-life}

\vspace{-0.1in}
\subsection{Infectious Disease Modeling}
We integrate an agent-based infectious disease model into the patterns-of-life simulation to capture individual-level transmission, disease progression, and pathogen shedding.
Each agent follows a cyclic SEIR process:
\[
S \rightarrow E \rightarrow I \rightarrow R \rightarrow S.
\]
Here, $S$ denotes susceptible, $E$ exposed, $I$ infectious, and $R$ recovered. Susceptible agents may become exposed through contact with infectious agents. Exposed agents are infected but not yet infectious. After the latent period, agents become infectious, enabling disease transmission and pathogen shedding. Following the infectious period, agents recover and acquire temporary immunity before eventually returning to the susceptible state as immunity wanes.

Transmission is probabilistic and depends on multiple conditions, including agent-level transmissibility, recipient susceptibility, a global infection ratio, and a cap on the number of secondary infections per agent, producing stochastic yet controlled disease spread.
In other words, when a susceptible agent encounters an infectious agent, transmission occurs only if the following independent conditions are satisfied: 
\[
u_1 < p_{\text{spread}}, \quad
u_2 < p_{\text{infect}}, \quad
u_3 < \rho, \quad
\text{and } n_{\text{spread}} < N_{\max},
\]
where \(u_1, u_2, u_3 \sim U(0,1)\) are independent uniform random samples, \(p_{\text{spread}}\) denotes the transmissibility of the infectious agent, \(p_{\text{infect}}\) the susceptibility of the recipient, \(\rho\) a global infection ratio regulating baseline transmissibility, \(n_{\text{spread}}\) the number of times the agent has attempted to spread the disease, and \(N_{\max}\) the maximum number of infections allowed per infectious agent. Only when all conditions hold does the susceptible agent transition to the exposed state.
Equivalently, for a meeting between infectious source \(y\) and susceptible recipient \(z\), infection occurs according to
\[
\Pr(z\ \text{becomes } E \mid y\ \text{meets } z)
= \mathbf{1}\!\{n_{\text{spread}} < N_{\max}\}
\times p_{\text{spread}} \times p_{\text{infect}} \times \rho.
\]
Each contact event is treated as independent, and a successful transmission \textbf{records the source agent as the infector} and transitions the recipient to the exposed state.

After infection, each agent \(a\) maintains individualized durations for disease stages: exposed, infectious, and recovered, denoted by \(\theta_E(a)\), \(\theta_I(a)\), and \(\theta_R(a)\). Let \(d_E(a)\), \(d_I(a)\), and \(d_R(a)\) represent the elapsed time in each state. Transitions occur as follows:
\[
\begin{aligned}
E \rightarrow I &\quad \text{if } d_E(a) \geq \theta_E(a), \\
I \rightarrow R &\quad \text{if } d_I(a) \geq \theta_I(a), \\
R \rightarrow S &\quad \text{if } d_R(a) \geq \theta_R(a).
\end{aligned}
\]
Upon returning to the susceptible state, infection counters and identifiers are reset so that future infections are treated independently.

Infection occurs through three interaction modes: incidental co-location, social encounters with new contacts, and repeated interactions within close social networks, capturing both spatial proximity and social structure.
To avoid synchronized transitions and better reflect real population variability, heterogeneity is introduced via a smoothness parameter \(s \sim U(0,1)\). Given global base durations \(\bar{\theta}_E\), \(\bar{\theta}_I\), and \(\bar{\theta}_R\), each agent receives adjusted durations computed as
\[
\tilde{\theta} = \left\lceil \bar{\theta} + (0.5 - s)\bar{\theta} \right\rceil.
\]
This stochastic perturbation spreads transitions across time, producing smoother epidemic dynamics~\cite{lloyd2001realistic}.

\vspace{-0.1in}
\subsection{Pathogen Shedding Dynamics}

Each infected agent begins shedding pathogens after a latent period following infection. In this simulation, the onset of shedding coincides with the period during which the agent is considered infectious. Pathogen shedding over time is modeled using a gamma-like function that captures the typical rise and subsequent decline in infectiousness observed in many infections.

The number of pathogens shed \(t\) days after the agent becomes exposed is given by
\begin{equation}
N(t) = N_0\, t^{b} e^{-a t},
\label{equ:pathogen}
\end{equation}
where \(N_0\) is a scaling constant representing the initial shedding magnitude at the onset of infectiousness, while \(a\) and \(b\) are rate and shape parameters controlling the decay speed and timing of peak shedding, respectively. Larger values of \(b\) delay the shedding peak, whereas larger values of \(a\) lead to faster decline after the peak.

The infection model operates in discrete time, where each simulation tick represents one unit of simulated time (e.g., 5 minutes). During each tick, every agent independently updates its internal disease state and interacts with other agents, thereby coupling biological progression with social behavior.
At each step, an agent increments the time spent in its current health state (\(d_E\), \(d_I\), or \(d_R\)), updates its pathogen load if it is in the infectious state, and evaluates state-transition conditions based on the corresponding thresholds \(\theta_E\), \(\theta_I\), and \(\theta_R\). When a threshold is reached, the agent transitions to the next disease state according to the model rules.
This iterative update process allows infection dynamics to evolve naturally as agents move, interact, and change health states throughout the simulation.

\begin{table}[b]
\centering
\caption{Simulation parameters and values. Names are shortened and written in a readable format. Complete names can be found in the \texttt{modified.properties} file for each instance.}
\label{tab:params}
\small
\begin{tabular}{|p{0.5\linewidth} |  p{0.4\linewidth} |}
\hline
\textbf{Parameter} & \textbf{Values} \\ \hline
Number of Agents & 1000, 10,000 \\  \hline
Latent Period (days) & 7 \\ \hline
Infectious Period (days) & 14 \\  \hline
Immunity Period (days) & 90, 2800 \\  \hline
Transmission Attempts per Agent & 50 \\  \hline
Initial Infected Agents & 10 \\ \hline
Transmission per Building per Agent & 10 \\ \hline
Work Hours per Day & 4, 8, 12 \\ \hline
Infection Rate ($\rho$) & 0.1, 0.15, 0.2, 0.25, 0.3, 0.5, 0.9 \\ \hline
\end{tabular}
\end{table}

\begin{table*}[h]
\centering
\caption{Pathogen Shedding Events  (\textit{poop\_in.parquet})}
\label{tab:shedding_data}
\small
\begin{tabular}{| C{0.06\linewidth} | C{0.12\linewidth} | C{0.1\linewidth} | C{0.1\linewidth} | C{0.08\linewidth} | C{0.1\linewidth} | C{0.1\linewidth} | C{0.12\linewidth} |}
\hline
agent\_id & time & latitude & longitude &venue\_type & pathogen\_level & disease\_status & infectious\_time \\ \hline
99 & 2024-01-01 04:35 & 34.090747 & -84.272987 & Workplace & 0 & Susceptible & \\ \hline
343 & 2024-01-01 04:40 & 33.747688  & -84.366316  & Workplace & 0 & Exposed & \\ \hline
4 & 2024-01-04 23:55 & 34.033755  & -84.180945  & Apartment & 18625732.6  & Infectious & 2024-01-01 00:00 \\ \hline
4 & 2024-01-05 04:55 & 33.747688  & -84.366316  & Workplace & 25178757.4  & Infectious & 2024-01-01 00:00 \\ \hline
647 & 2024-01-14 05:05 & 33.616730  & -84.540017  & Restaurant & 4927948.9  & Infectious & 2024-01-12 05:05 \\ \hline

\end{tabular}
\end{table*}

\begin{table*}[h]
\centering
\caption{Disease Status  (\textit{disease\_status.parquet})}
\label{tab:disease_data}
\small
\begin{tabular}{ | C{0.11\linewidth} | C{0.05\linewidth} | C{0.11\linewidth} | C{0.11\linewidth} | C{0.07\linewidth} | C{0.09\linewidth} | C{0.05\linewidth} |  C{0.08\linewidth} |  C{0.1\linewidth} |}
\hline
time & agent\_id & exposed\_time & infectious\_time & pathogen & disease\_status & source & latitude & longitude \\
\hline
2024-01-01 00:05 & 202 &  &  & 0 & Susceptible & -1 & 33.750634  & -84.416203 \\ \hline
2024-02-09 00:00 & 202 & 2024-02-08 09:45 &  & 0 & Exposed & 413 & 33.750634  & -84.416203 \\ \hline
2024-02-15 00:00 & 202 & 2024-02-08 09:45 & 2024-02-13 09:45 & 541091.1 & Infectious & 413 & 33.750634  & -84.416203 \\ \hline
2024-02-18 00:55 & 202 & 2024-02-08 09:45 & 2024-02-13 09:45 & 87898892.1 & Infectious & 413 & 33.750634  & -84.416203 \\ \hline
2024-02-26 00:55 & 202 & 2024-02-08 09:45 & 2024-02-13 09:45 & 64899.5 & Recovered & 413 & 33.750634  & -84.416203 \\ \hline
\end{tabular}
\end{table*}

\vspace{-0.1in}
\subsection{Defecation Behavior}

We extend the patterns-of-life simulation by incorporating a physiological defecation mechanism that models the gradual buildup and relief of a biological need over time, enabling more realistic behavior and supporting wastewater-based epidemiology analyses. This simulated buildup is modeled after real-world phases and transitions that occur in defecation \cite{heitmann2021understanding}. Each agent maintains an internal defecation state that evolves continuously according to an individual defecation rate sampled from a uniform distribution. Agents transition through five states: \textit{Just Defecated}, \textit{No Need To Defecate}, \textit{Building Pressure}, \textit{Need To Defecate}, and \textit{Urgent Need To Defecate}. State transitions depend on elapsed time since the last event and threshold-based pressure levels. After relief, the internal state resets and a new cycle begins. Agents may satisfy this need only when not in transit and when located in suitable environments such as home, work, or recreational sites. As urgency grows, agents may temporarily reprioritize activities to satisfy the need, generating realistic variation in daily routines, mobility patterns, and waste production across the simulated population. 

The duration and frequency of defecation in the human population varies widely, and while population surveys have shed light on some aspects of patterns of shedding in humans, much remains unknown \cite{mitsuhashi2018characterizing}.  To control the duration of comfort periods while allowing for variation, we use a defecation rate, \(\eta\), drawn from a uniform distribution that determines the speed of pressure buildup and the urgency thresholds. After each defecation event, agents experience a no-urge period lasting \(60 - 30\eta\) minutes, followed by a comfort period of \(180 - 60\eta\) minutes during which internal pressure gradually increases. The internal level then decreases at a rate proportional to \(0.65\eta\). When this level falls below \(30 + 20\eta\), the agent enters the need state, and reaching zero corresponds to the urgent state. After relief, the internal state resets, allowing multiple daily events depending on individual physiological rates.

\section{Experimental Results}
\label{sec:results}
In this section, we describe the configuration of the simulation parameters and the generated simulation outputs. Using different parameter configurations, we execute the simulation framework under multiple scenarios and analyze the resulting mobility, epidemiological, and wastewater-related behaviors.
\vspace{-0.1in}

\subsection{Environmental Setup}
The experiments were executed on a machine equipped with an Intel Core i7-6700HQ CPU, 24~GB of RAM, and running Fedora Linux 43. Simulation parameters were varied to evaluate the impact of different scenarios on mobility and behavioral outcomes. \reftab{tab:params} summarizes the primary simulation parameters and their configured values.
All remaining parameters use the default values from~\cite{amiri2026we}. 

In addition, the pathogen control parameters in \refequ{equ:pathogen} are set to \(a = 2\), \(b = 8\), and \(N_0 = 10^7\). With this configuration, pathogen shedding rises sharply and then decays gradually, producing a right-skewed bell-shaped curve that is consistent with typical infectiousness profiles observed in~\cite{he2020temporal}.

Using these configurations, the simulation framework produced 84 instances of the simulation world under different epidemiological and behavioral settings.
For each instance, the framework generates datasets covering:
Pathogen Shedding Events (\textit{poop\_in.parquet});
Disease Status Records (\textit{disease\_status.parquet});
Mobility Check-in Records (\textit{check\_in.parquet});
Social Interaction Network data (\textit{social\_links.parquet});
and Agent Home Locations (\textit{home\_locations.csv}).

Together, these outputs provide a synthetic benchmark environment for mobility-aware wastewater epidemiology analysis and downstream spatial computing tasks such as outbreak monitoring, hotspot analysis, and surveillance strategy evaluation.
\vspace{-0.1in}

\subsection{Simulation Results}
The simulation framework produces multiple datasets capturing agent mobility, interaction patterns, infection dynamics, and pathogen shedding events. Each dataset is stored in tabular format and can be linked using the common \texttt{agent\_id}. The main data products are summarized in this section, with additional details provided in the accompanying data documentation. For readability, some column names and values are simplified here; users should refer to the released datasets for the complete names and values.
\subsubsection{Pathogen Shedding Events}
Table~\ref{tab:shedding_data} records pathogen shedding events occurring at different venues and times. Each entry includes the agent identifier, event timestamp, event location, venue type, pathogen level, disease status, and the timestamp when the agent transitioned to infectious status. These events capture environmental contamination resulting from agent presence and support downstream wastewater epidemiology applications such as hotspot analysis, outbreak monitoring, and spatial transmission inference.

\subsubsection{Disease Status Records}
Table~\ref{tab:disease_data} records the temporal progression of infection states for agents. Each entry includes the observation time, agent identifier, timestamps marking transitions between disease stages, pathogen load, and the current disease status. Additional fields provide the infection source identifier and the agent’s location at the time of observation. This dataset supports reconstruction of transmission chains, benchmarking of epidemiological models, and analysis of outbreak dynamics over time.

\begin{table}[h]
\centering

\vspace{-0.1in}
\caption{Social Interaction Network  (\textit{social\_links.parquet})}

\vspace{-0.1in}
\label{tab:social_data}
\small
\begin{tabular}{|C{0.3\linewidth} | C{0.12\linewidth} | C{0.12\linewidth} |}
\hline
time & from & to \\ \hline
2024-01-01 00:00:00 & 670 & 684 \\ \hline 
2024-01-01 00:00:00 & 670 & 677 \\ \hline 
2024-01-01 00:00:00 & 670 & 680 \\ \hline 
\end{tabular}

\vspace{-0.1in}
\end{table}

\subsubsection{Social Interaction Network}
Table~\ref{tab:social_data} represents directed social interactions between agents at given timestamps. Each record defines an interaction edge from one agent to another, allowing construction of dynamic social graphs. These networks support analyses of contact clustering, social connectivity, and mobility-aware disease propagation throughout the simulation.

\begin{table}[h]
\centering
\caption{Mobility Check-in  (\textit{check\_in.parquet})}
\label{tab:checkin_data}
\small
\begin{tabular} {| C{0.06\linewidth} | C{0.15\linewidth} | C{0.07\linewidth} |C{0.15\linewidth} | C{0.12\linewidth} |C{0.15\linewidth} |}
\hline
agent & time & venue & venue\_type & latitude & longitude \\
\hline
0 & 2024-01-01 12:25 & 1754 & Pub & 33.733581  & -84.414484 \\ \hline
65 & 2024-02-11 06:20 & 1770 & Restaurant & 33.896081  & -84.382190 \\ \hline
519 & 2024-04-25 08:25 & 1733 & Workplace & 34.022058  & -84.286243 \\ \hline
999 & 2024-09-07 02:30 & 600 & Apartment & 34.085193 & -84.350052 \\ \hline
\end{tabular}
\end{table}

\subsubsection{Mobility Check-in Records}
Table~\ref{tab:checkin_data} records agent mobility in the form of time-stamped venue check-ins. Each entry includes agent identifier, visit time, venue identifier, venue category, and geographic coordinates. These records allow reconstruction of movement trajectories and identification of potential contact opportunities between agents. Venue types include workplaces, residences, and public locations, enabling applications in spatial epidemiology, mobility analysis, and wastewater-based surveillance benchmarking.

\subsection{Visualization of Results}
\begin{figure*}[h]
    \centering
    \begin{minipage}{0.33\textwidth}
        \centering
        \includegraphics[width=\linewidth]{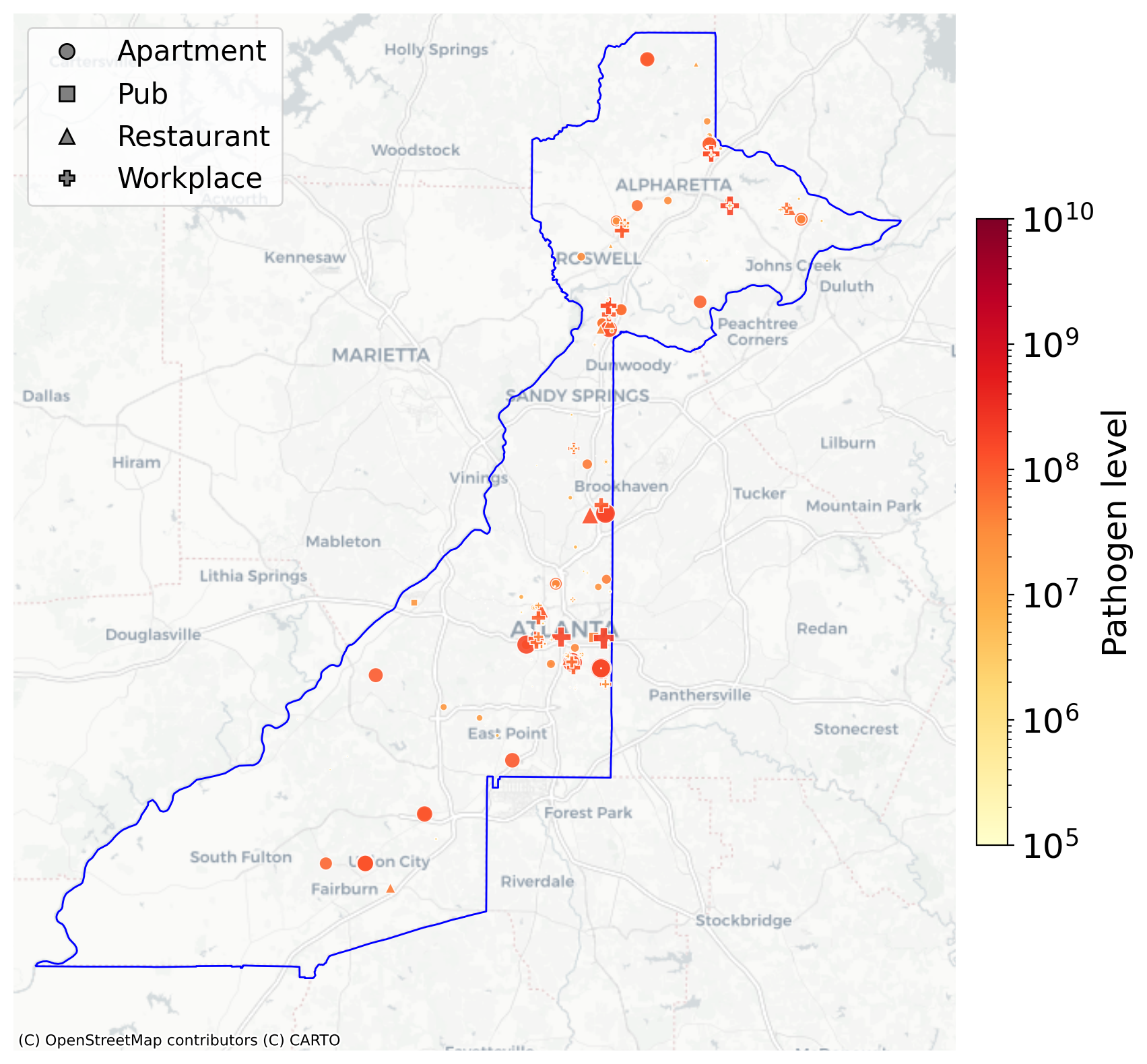}
        
        \subcaption{February 1, 2024}
    \end{minipage}\hfill
    \begin{minipage}{0.33\textwidth}
        \centering
        \includegraphics[width=\linewidth]{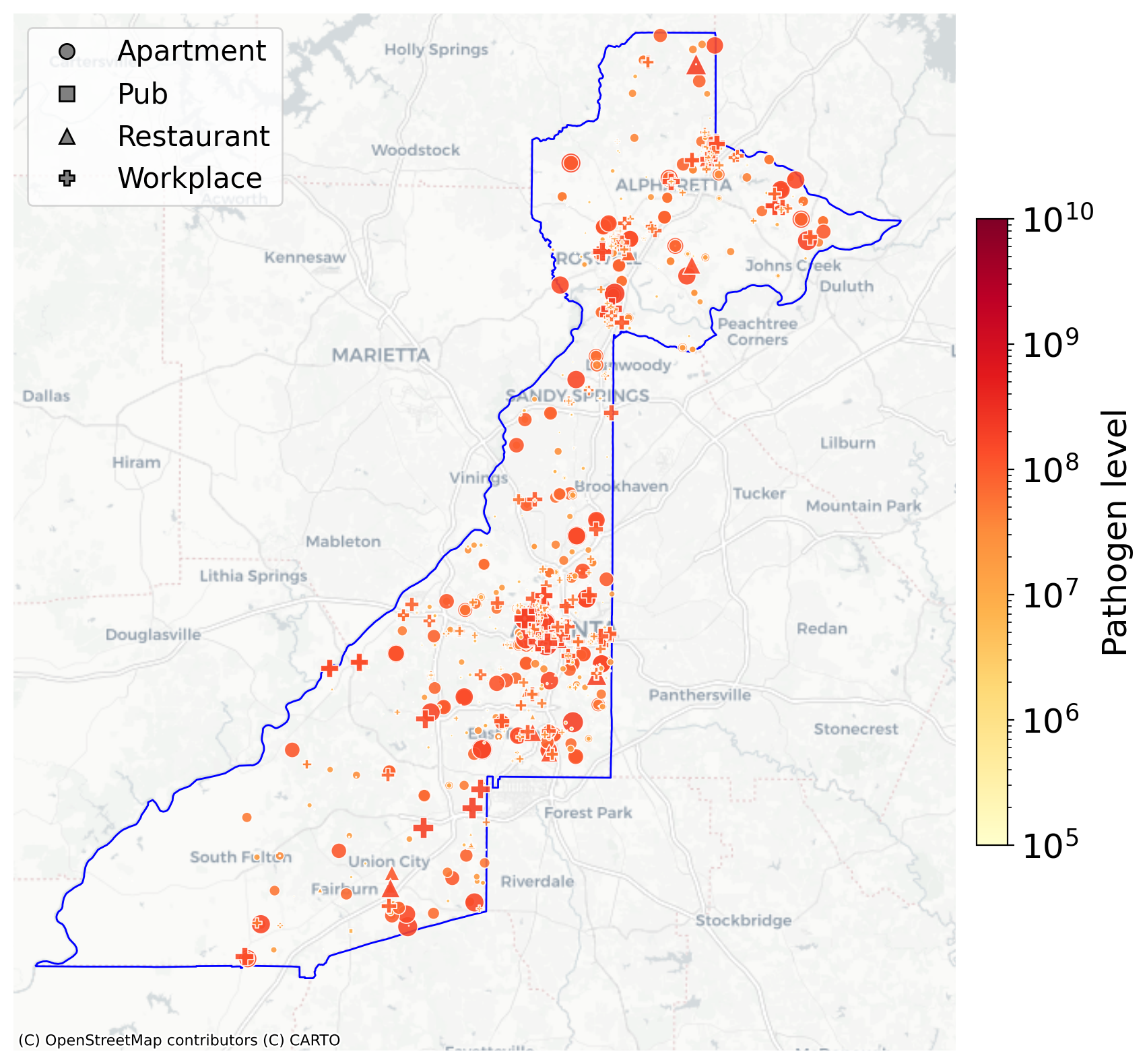}
        
        \subcaption{March 1, 2024}
    \end{minipage}\hfill
    \begin{minipage}{0.33\textwidth}
        \centering
        \includegraphics[width=\linewidth]{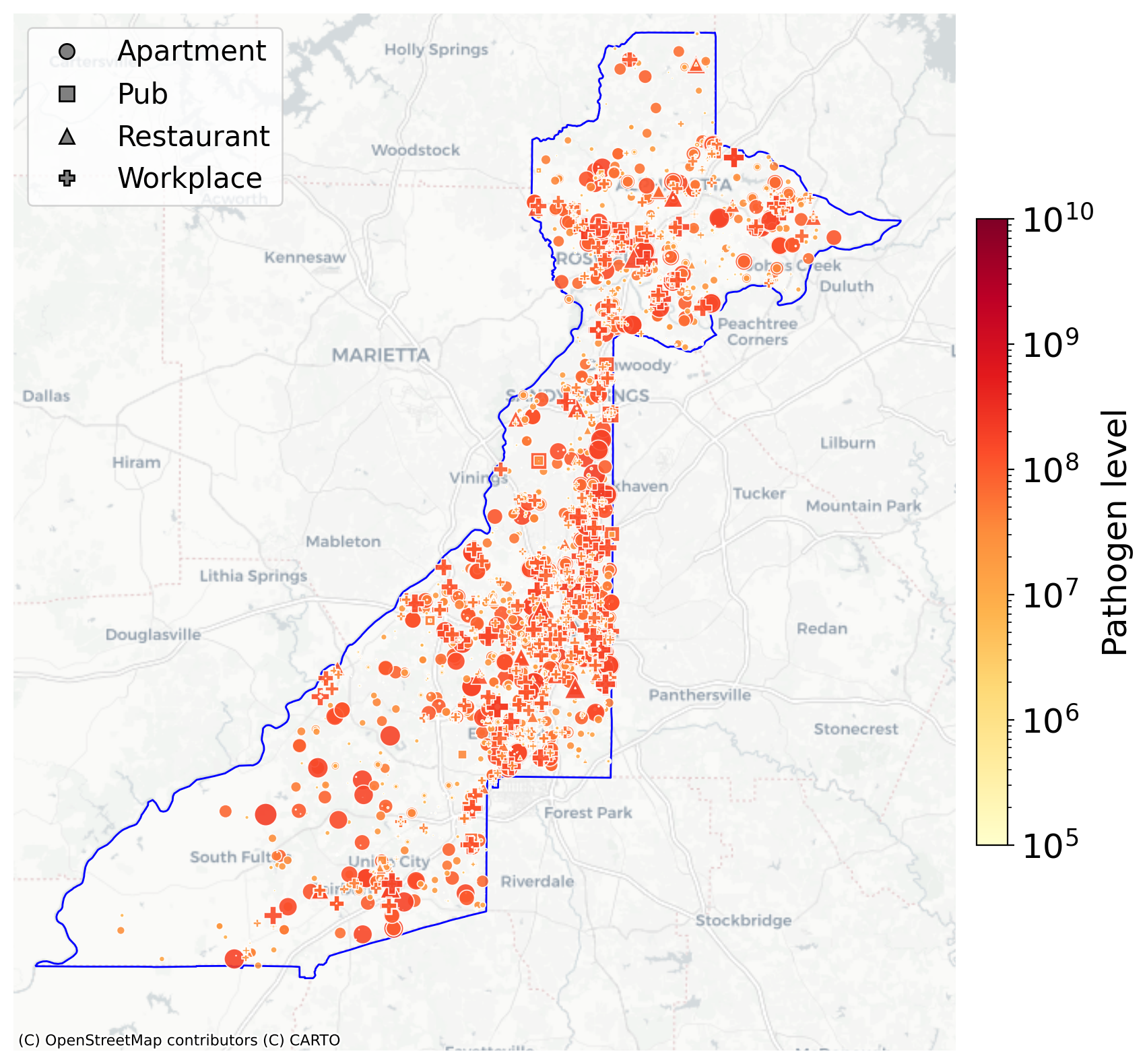}
        
        \subcaption{April 1, 2024}
    \end{minipage}
    \vspace{-0.1in}
    \caption{Spatial distribution of pathogen load over the study area for a representative scenario. Points indicate sampled venues (colored/marked by venue type), and the color scale represents pathogen level at those locations (population $=10{,}000$, $\rho=0.3$, immunity $=90$ days, work hours/day $=8$)}
    \label{fig:spatial_pathogen_maps}
     \vspace{-0.1in}
\end{figure*}

\begin{figure*}[h]
    \centering
    \begin{minipage}{0.29\textwidth}
        \centering
        \includegraphics[width=\linewidth, trim=0 0 240 0, clip]{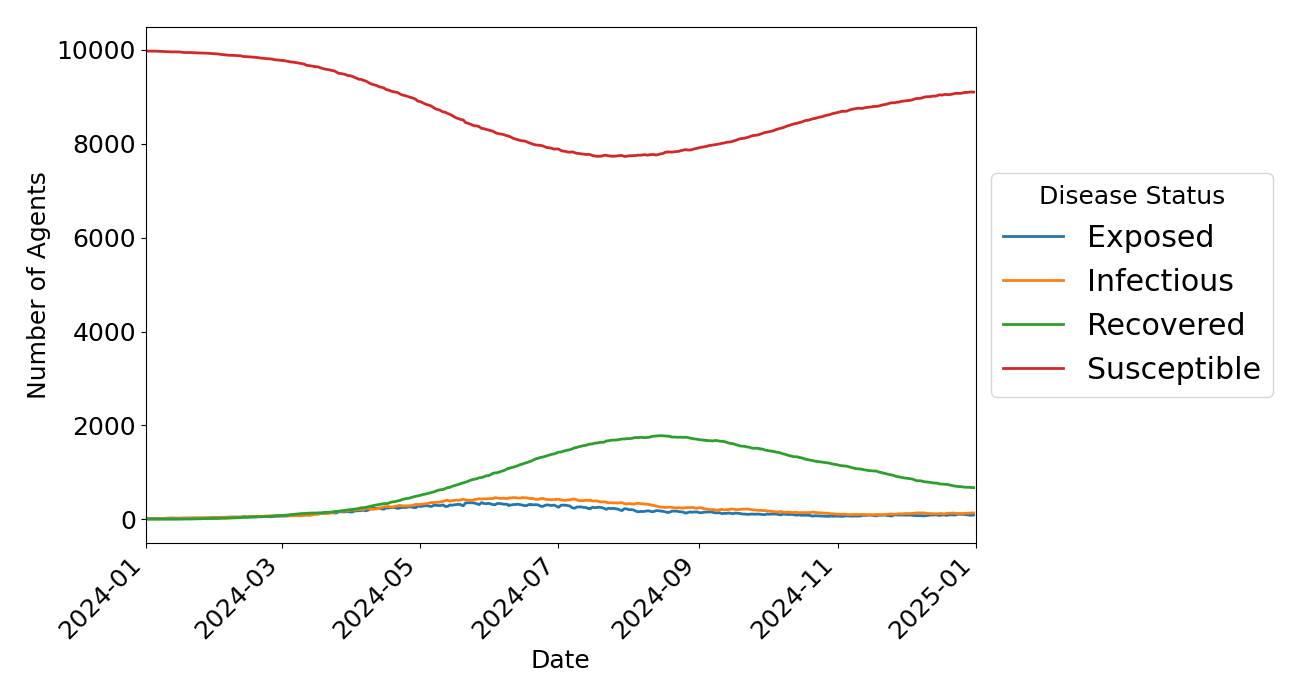}
        
        \subcaption{$\rho=0.15$}
    \end{minipage}\hfill
    \begin{minipage}{0.29\textwidth}
        \centering
        \includegraphics[width=\linewidth, trim=0 0 240 0, clip]{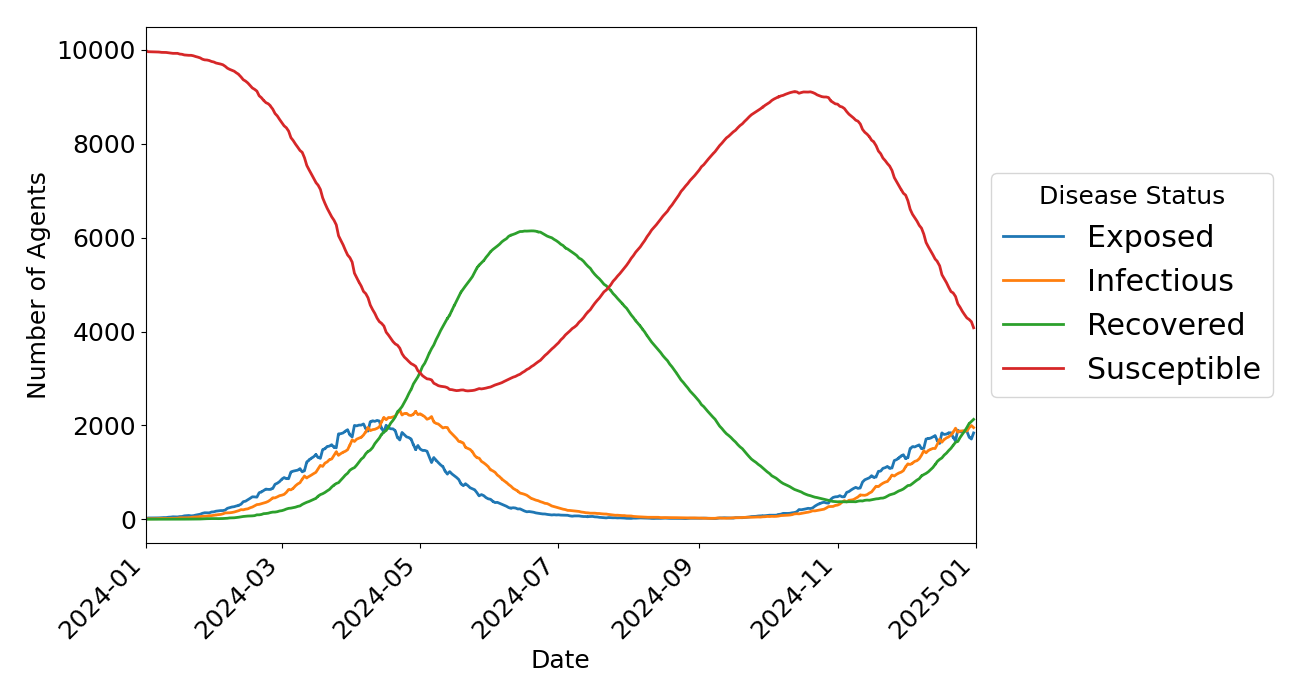}
        
        \subcaption{$\rho=0.3$}
    \end{minipage}\hfill
    \begin{minipage}{0.29\textwidth}
        \centering
        \includegraphics[width=\linewidth, trim=0 0 240 0, clip]{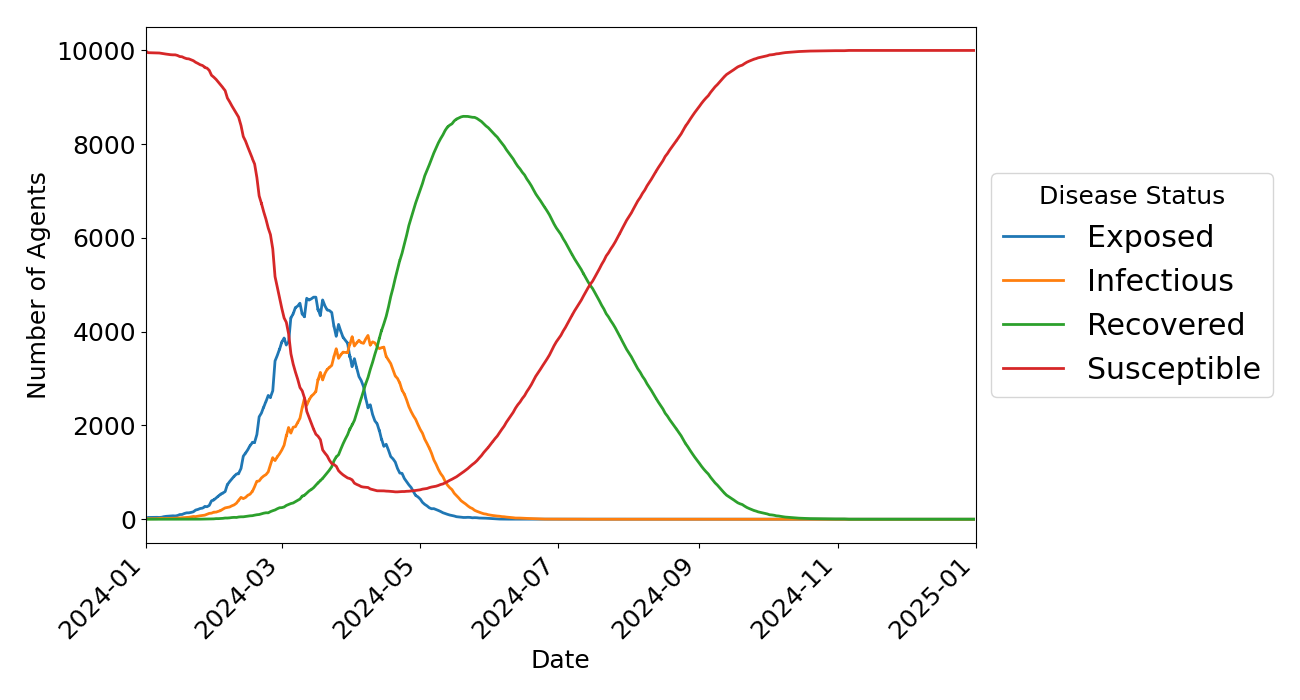}
        
        \subcaption{$\rho=0.9$}
    \end{minipage}\hfill
     \begin{minipage}{0.1\textwidth}
        \centering
        \includegraphics[width=\linewidth, trim=710 0 0 0, clip]{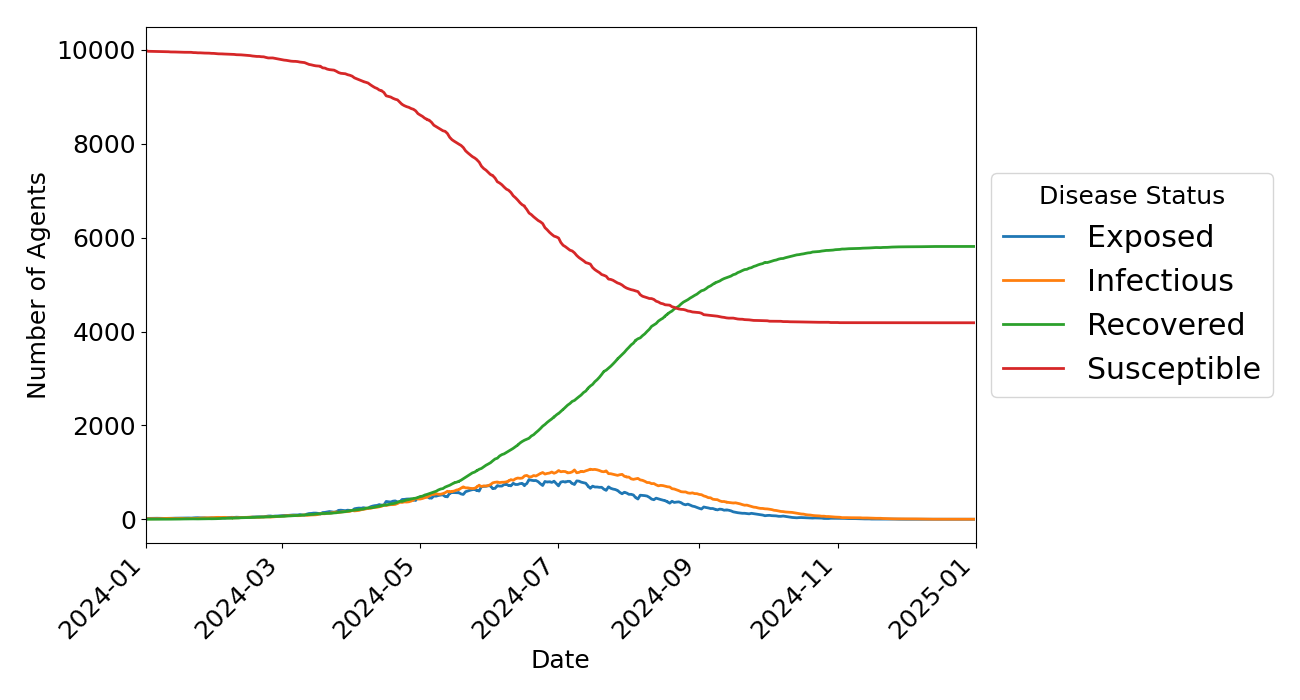}
    \end{minipage}
     \vspace{-0.1in}
    \caption{Epidemiological curve for different infection rates $\rho$ (population $=10{,}000$, immunity $=90$ days, work hours/day $=8$). \vspace{-0.05in}} \vspace{-0.1in}
    \label{fig:seir_vary_ir}
\end{figure*}

\begin{figure*}[h]
    \centering
    \begin{minipage}{0.33\textwidth}
        \centering
        \includegraphics[width=\linewidth]{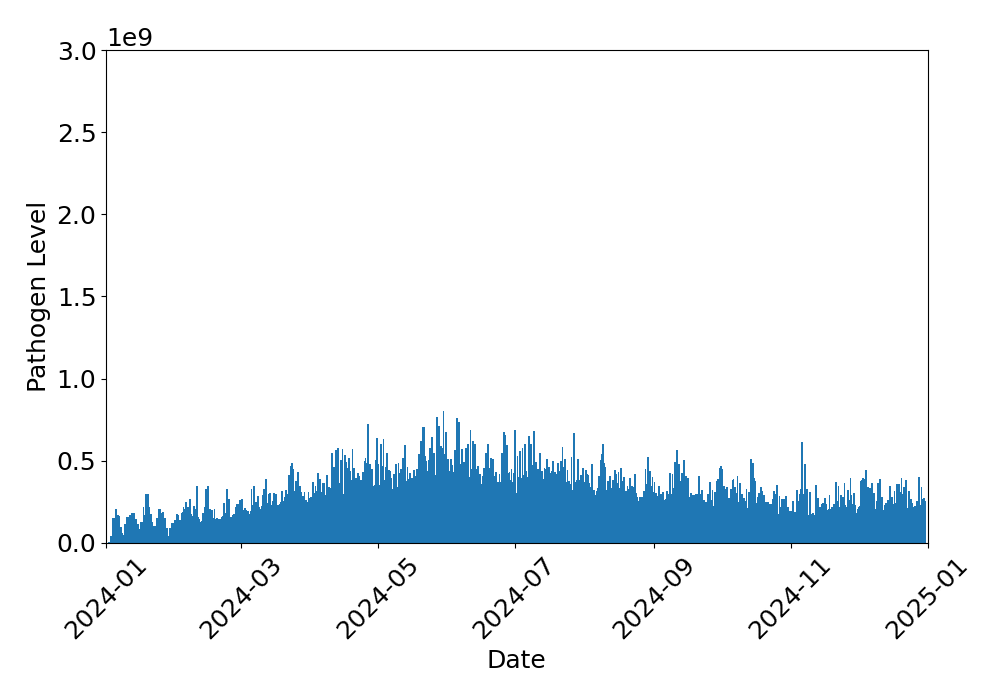}
        
        \subcaption{$\rho=0.15$}
    \end{minipage}\hfill
    \begin{minipage}{0.33\textwidth}
        \centering
        \includegraphics[width=\linewidth]{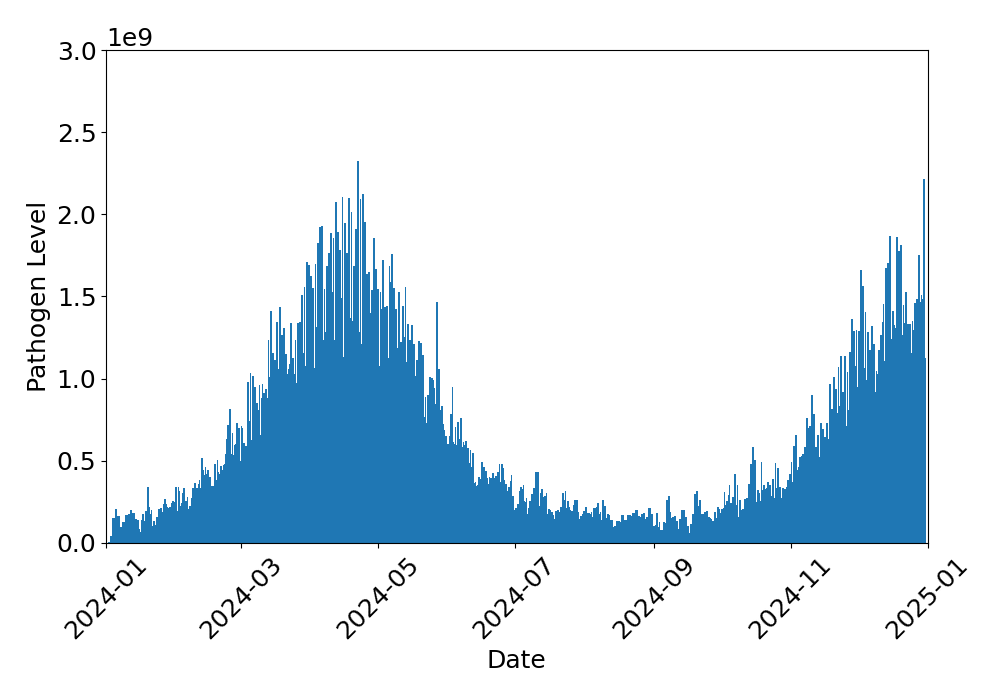}
        
        \subcaption{$\rho=0.3$}
    \end{minipage}\hfill
    \begin{minipage}{0.33\textwidth}
        \centering
        \includegraphics[width=\linewidth]{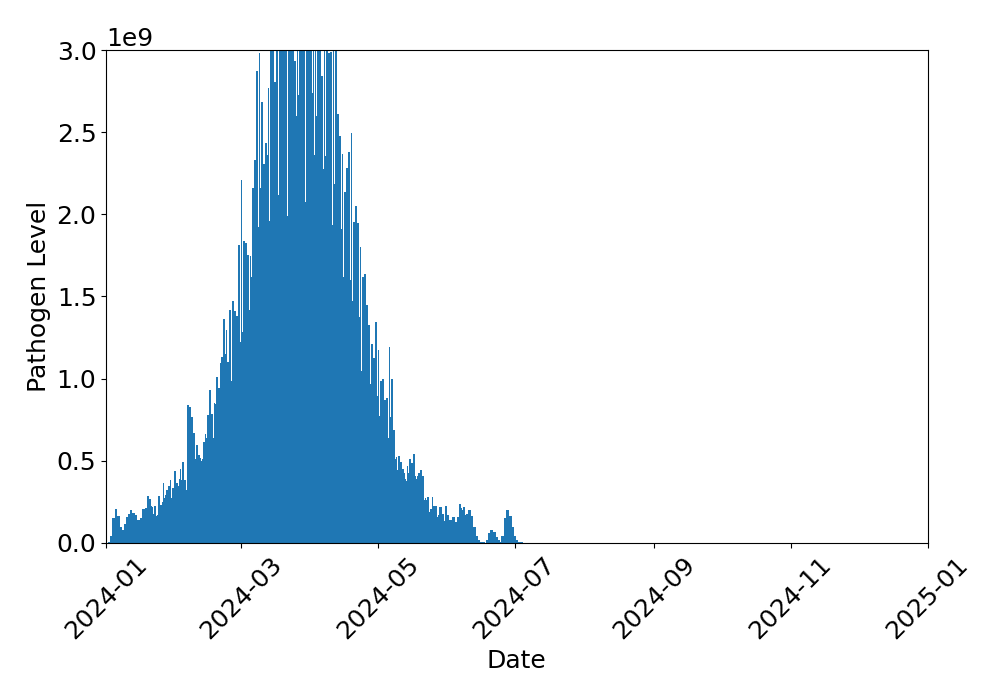}
        
        \subcaption{$\rho=0.9$}
    \end{minipage}
     \vspace{-0.1in}
    \caption{Aggregated wastewater pathogen signal for different infection rates $\rho$ (same settings as Figure~\ref{fig:seir_vary_ir}). \vspace{-0.05in}} \vspace{-0.1in}
    \label{fig:pathogen_vary_ir}
\end{figure*}

\begin{figure*}[h]
    \centering
    \begin{minipage}{0.29\textwidth}
        \centering
        \includegraphics[width=\linewidth, trim=0 0 240 0, clip]{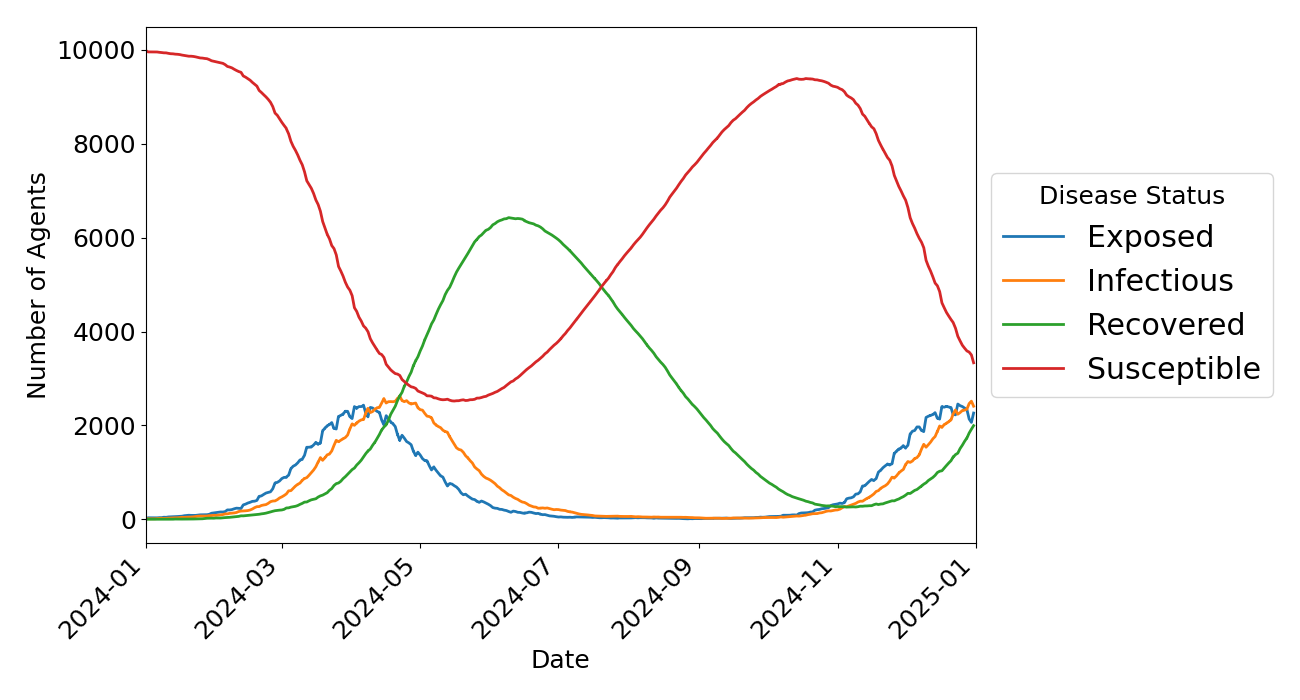}
        
        \subcaption{4 hours/day}
    \end{minipage}\hfill
    \begin{minipage}{0.29\textwidth}
        \centering
        \includegraphics[width=\linewidth, trim=0 0 240 0, clip]{figs/disease/out/pop_10000_ir_0.3_rdy_90_whpd_8.png}
        
        \subcaption{8 hours/day}
    \end{minipage}\hfill
    \begin{minipage}{0.29\textwidth}
        \centering
        \includegraphics[width=\linewidth, trim=0 0 240 0, clip]{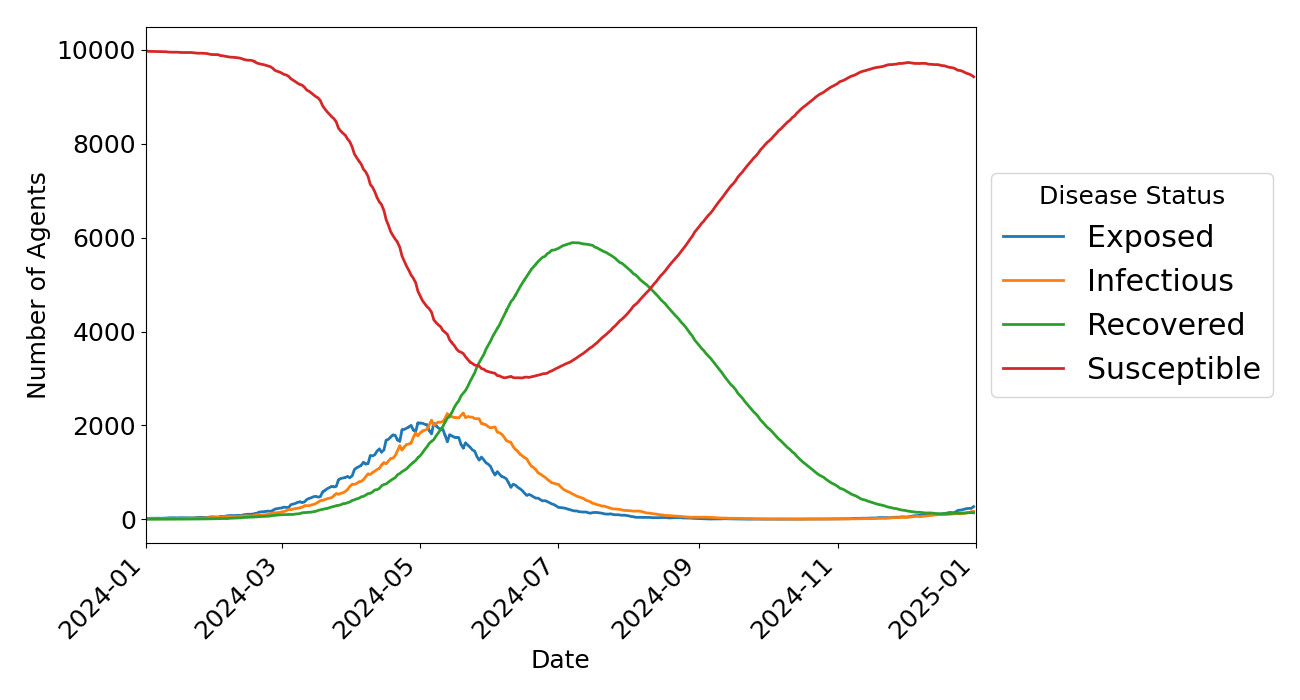}
        
        \subcaption{12 hours/day}
    \end{minipage}\hfill
     \begin{minipage}{0.1\textwidth}
        \centering
        \includegraphics[width=\linewidth, trim=710 0 0 0, clip]{figs/disease/out/pop_10000_ir_0.2_rdy_2800_whpd_12.png}
    \end{minipage}
    \vspace{-0.1in}
    \caption{Epidemiological curve for different work hours/day (population $=10{,}000$, immunity $=90$ days, infection rate $\rho=0.3$).\vspace{-0.05in}}\vspace{-0.05in}
    \label{fig:seir_vary_workhours}
\end{figure*}

\begin{figure*}[h]
    \centering
    \begin{minipage}{0.33\textwidth}
        \centering
        \includegraphics[width=\linewidth]{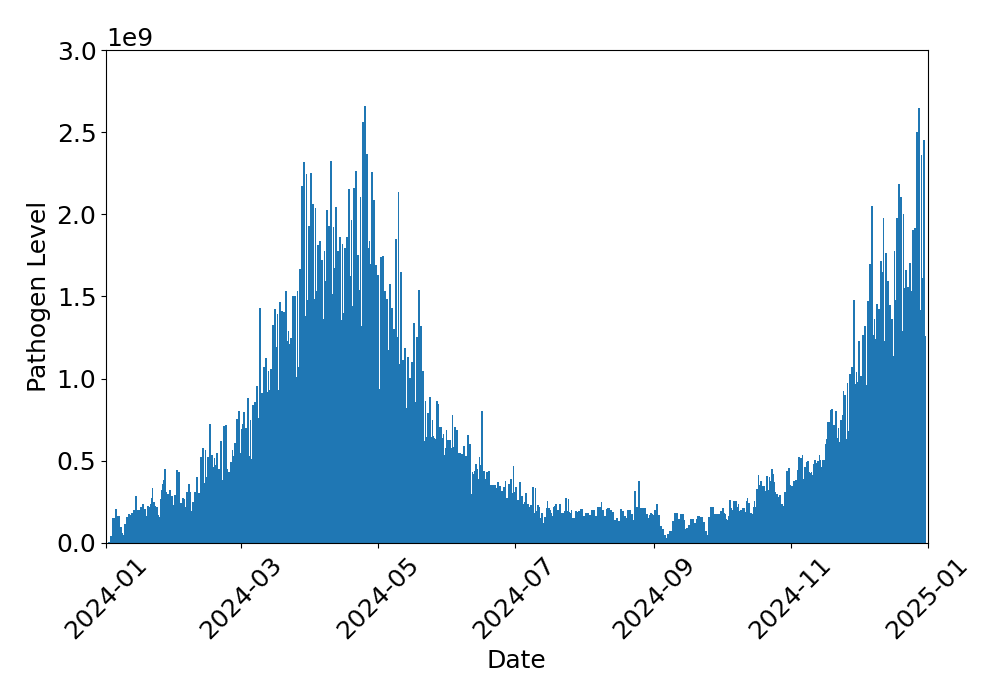}
        
        \subcaption{4 hours/day}
    \end{minipage}\hfill
    \begin{minipage}{0.33\textwidth}
        \centering
        \includegraphics[width=\linewidth]{figs/pathogen/pop_10000_ir_0.3_rdy_90_whpd_8.png}
        
        \subcaption{8 hours/day}
    \end{minipage}\hfill
    \begin{minipage}{0.33\textwidth}
        \centering
        \includegraphics[width=\linewidth]{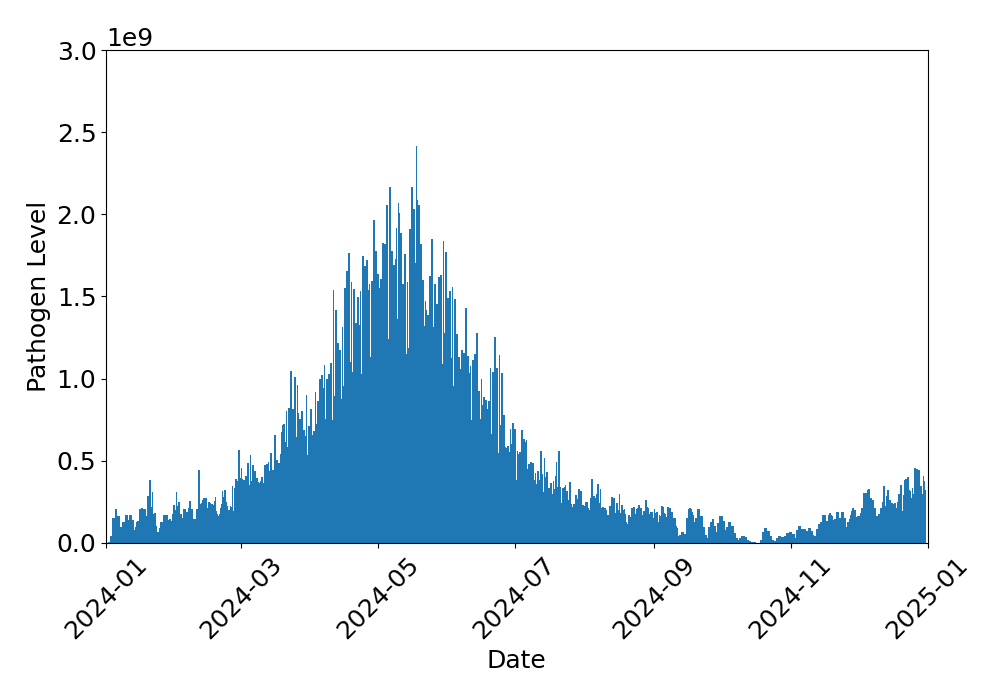}
        
        \subcaption{12 hours/day}
    \end{minipage}
    \vspace{-0.1in}
    \caption{Aggregated wastewater pathogen signal under different work hours/day (same settings as Figure~\ref{fig:seir_vary_workhours}).\vspace{-0.05in}}\vspace{-0.05in}
    \label{fig:pathogen_vary_workhours}
\end{figure*}

We visualize simulated outbreak dynamics and the corresponding wastewater signals under multiple scenario settings. Across plots, we report: 
(i) epidemiological curve, the number of agents in each SEIR state over time (Susceptible, Exposed, Infectious, Recovered), and (ii) aggregated pathogen signals derived from individual shedding events recorded in \texttt{poop\_in.parquet}. All scenarios share the same base configuration (Table~\ref{tab:params}) and differ only in the parameter highlighted in each subsection. We visualize only a subset of the 84 generated datasets; all visualizations are provided in a separate folder alongside the data.

\vspace{-0.1in}
\subsubsection{Spatial analysis of wastewater pathogen load}
We analyze pathogen load across the sewer network to illustrate how infections and defecation behavior translate into geographically structured wastewater signals. \reffig{fig:spatial_pathogen_maps} shows the spatial distribution of pathogen load across the study area at three time points, February 1, March 1, and April 1, 2024, under a representative simulation scenario. Each point represents a sampled venue, including apartments, pubs, restaurants, and workplaces, while color intensity indicates the estimated pathogen level at each location. Over time, the number of affected venues increases and the distribution becomes more widespread, demonstrating progressive spatial spread and intensification of environmental contamination. By April, pathogen presence is dense across much of the region, reflecting sustained transmission and accumulation driven by agent mobility and interactions under the modeled conditions.

\begin{figure*}[t]
    \centering
    \begin{minipage}{0.45\textwidth}
        \centering
    \includegraphics[width=0.9\linewidth]{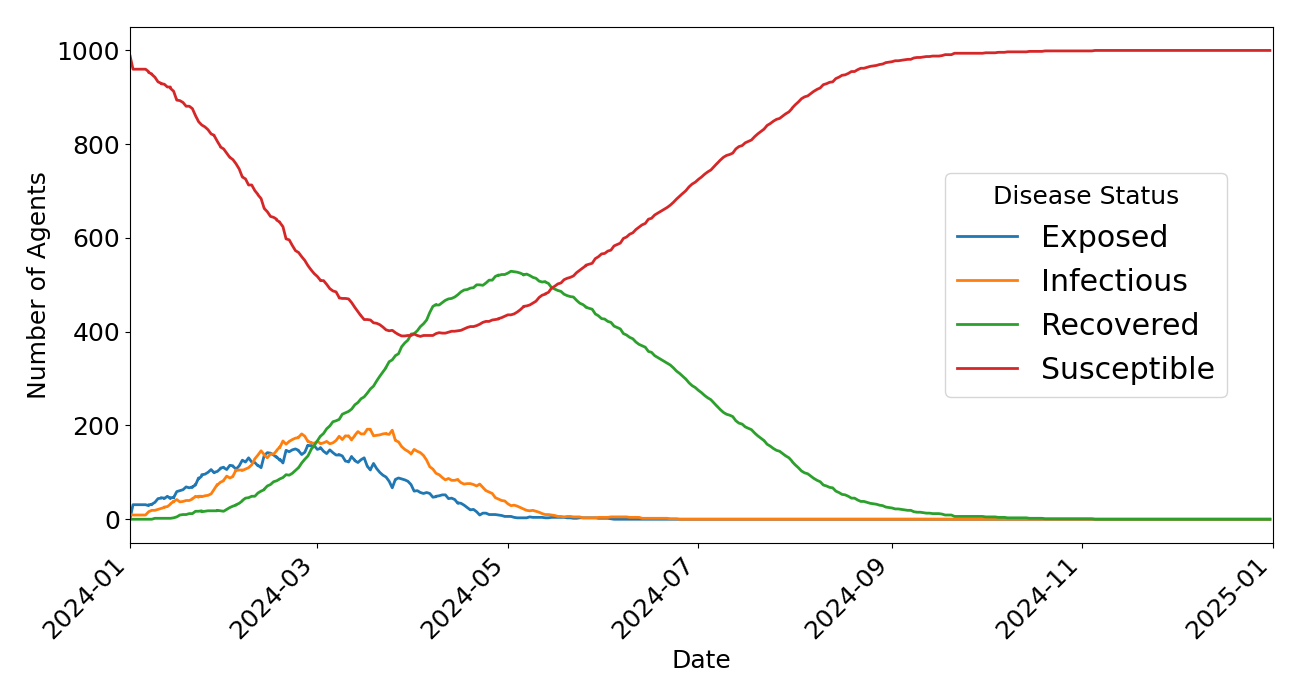}
    
        \subcaption{population $=1000$}
    \end{minipage}
    \begin{minipage}{0.45\textwidth}
        \centering
    \includegraphics[width=0.9\linewidth]{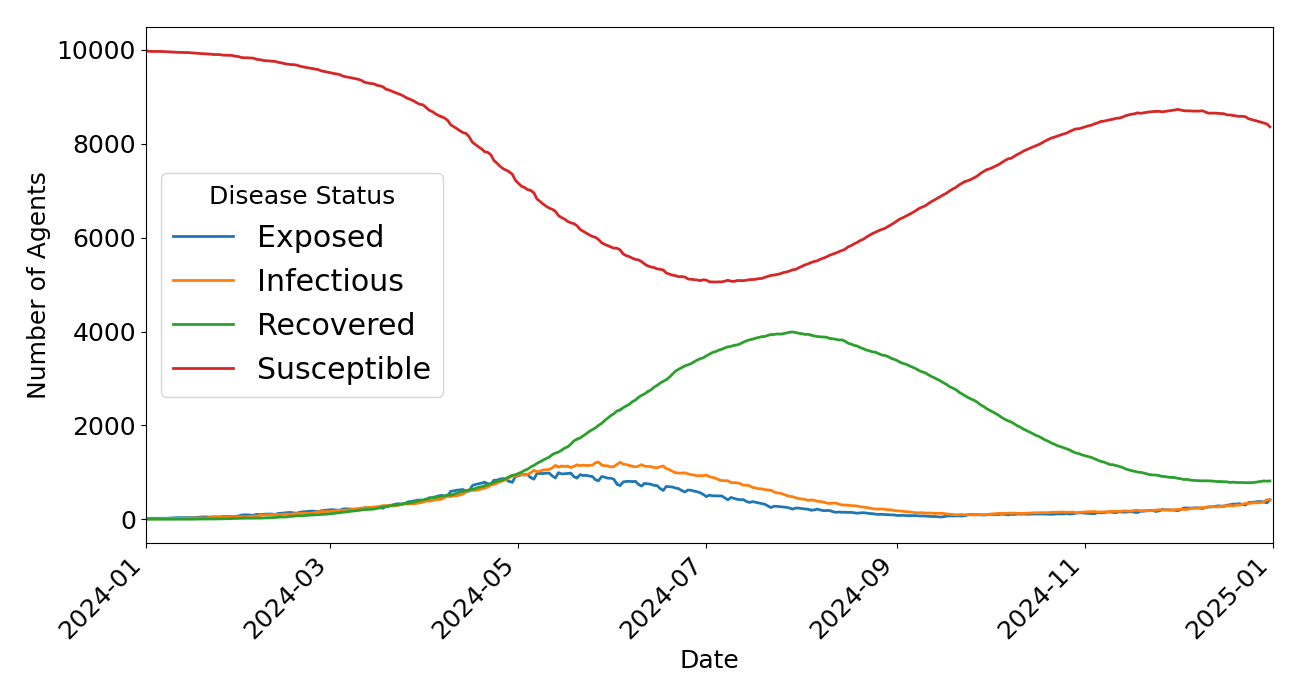}
    
        \subcaption{population $=10{,}000$}
    \end{minipage}
    \vspace{-0.1in}
    \caption{Epidemiological curve for different population (infection rate $\rho=0.2$, immunity $=90$ days, work hours/day $=8$).\vspace{-0.05in}}\vspace{-0.1in}
    \label{fig:seir_vary_pop}
\end{figure*}

\vspace{-0.1in}
\subsubsection{Varying infection rate}
We vary the global infection rate $\rho$ (Table~\ref{tab:params}) to examine how transmission intensity affects epidemic size and timing. Higher infection rates produce faster growth in the infectious population, earlier epidemic peaks, and more rapid depletion of susceptible agents. These dynamics also generate stronger and earlier wastewater pathogen signals.

\reffig{fig:seir_vary_ir} compares epidemiological curves under three infection rates, illustrating how transmission intensity influences outbreak dynamics in a population of 10,000 agents. Each panel shows the number of susceptible, exposed, infectious, and recovered individuals over time. At low transmission ($\rho = 0.15$), the outbreak grows slowly, leading to only a modest increase in infections while most of the population remains susceptible. With moderate transmission ($\rho = 0.3$), a pronounced epidemic wave emerges, producing large exposed and infectious populations followed by substantial recovery and a partial resurgence later in the year. Under high transmission ($\rho = 0.9$), infections rise rapidly and peak early, quickly depleting the susceptible population and moving most agents into the recovered class, after which transmission declines due to the limited number of remaining susceptible individuals. Overall, increasing infection rates accelerate spread, increase peak infections, and shorten epidemic duration.

\reffig{fig:pathogen_vary_ir} presents the aggregated wastewater pathogen signal over time for the same infection rates, showing how transmission intensity shapes environmental pathogen dynamics. At low transmission ($\rho = 0.15$), pathogen levels increase gradually and remain relatively low, reflecting limited infection spread and modest shedding into wastewater. For moderate transmission ($\rho = 0.3$), the signal shows a pronounced peak associated with a major epidemic wave, followed by a decline and a later resurgence as transmission continues within the population. Under high transmission ($\rho = 0.9$), pathogen levels rise rapidly to an early sharp peak and then decline quickly as susceptible individuals are depleted and transmission subsides. These results show that wastewater signals closely follow epidemic dynamics, with higher infection rates producing earlier and larger peaks followed by faster declines.

\vspace{-0.1in}
\subsubsection{Varying work hours per day}
We vary daily work hours to examine how routines and workplace exposure influence transmission dynamics. Longer workdays generally increase time spent in workplaces, creating more contact opportunities and altering the timing and magnitude of infection waves, and consequently the wastewater signal. In contrast, shorter work hours leave more time for social activities outside work, which can also contribute to transmission through increased community contacts.

\reffig{fig:seir_vary_workhours} compares epidemiological curves under different daily workplace exposure durations, illustrating how time spent at workplaces shapes transmission dynamics. With shorter workplace exposure (4 hours per day), infections grow more gradually and peak at lower levels, while a larger fraction of the population remains susceptible for a longer period. At the baseline setting (8 hours per day), a stronger epidemic wave develops, producing higher exposed and infectious populations and a substantial shift of agents into the recovered class. When workplace exposure increases to 12 hours per day, transmission accelerates further, leading to an earlier and more pronounced epidemic peak and faster depletion of susceptible individuals. Overall, increased time spent in workplaces intensifies transmission, advances epidemic timing, and increases the magnitude of infection waves.

\reffig{fig:pathogen_vary_workhours} shows the aggregated wastewater pathogen signal under different workplace exposure durations, demonstrating how time spent in shared work environments affects environmental pathogen levels. With shorter workplace exposure (4 hours per day), pathogen levels increase gradually and remain relatively moderate, reflecting slower transmission and lower overall shedding into wastewater. At the baseline exposure (8 hours per day), the signal exhibits a strong peak associated with a major epidemic wave, followed by a decline and a later resurgence as transmission persists within the population. When workplace exposure increases to 12 hours per day, pathogen levels rise more rapidly and reach a higher early peak, after which the signal declines as the susceptible population is depleted and transmission subsides. Overall, longer workplace exposure produces earlier and stronger wastewater signals, mirroring the acceleration and intensification of epidemic spread.

\vspace{-0.1in}
\subsubsection{Varying population size}
We compare outbreaks simulated with 1,000 and 10,000 agents to examine how population scale affects epidemic dynamics. As population size increases, absolute case counts grow while stochastic fluctuations become smaller relative to the total population, producing smoother epidemic trajectories.

\reffig{fig:seir_vary_pop} compares epidemiological curves for the two population sizes under identical transmission settings. In the smaller population (1,000 agents), infections rise quickly and peak earlier, rapidly reducing the susceptible population and producing a relatively short epidemic wave before transmission subsides. In contrast, the larger population (10,000 agents) exhibits a broader and more prolonged outbreak, with infections spreading more gradually and sustaining transmission over a longer period, resulting in a delayed peak and slower recovery dynamics. Overall, larger populations support longer and more sustained epidemic waves, while smaller populations experience faster but shorter outbreaks under comparable transmission conditions.

\section{Reproducibility and Data Availability}
\label{sec:reproducibility}
To reproduce the datasets and simulation scenarios presented in this study, users can run the simulation framework described in~\cite{amiri2026we}. The source code is available at \url{https://github.com/onspatial/wastewater-based-epidemiology-patterns-of-life}, where it can be compiled into a JAR file serving as the core simulation engine. This engine can then be used with the software available at \url{https://github.com/onspatial/hd-gen} to configure and execute multiple simulation instances in parallel, as described in~\cite{amiri2026hd}. This workflow enables reproducible generation of large-scale wastewater epidemiology datasets under diverse spatial, behavioral, and epidemiological scenarios.

To generate a single simulation instance or execute the framework manually, users can build the JAR file and run the provided \texttt{run.sh} script included with the dataset, modifying simulation parameters as needed. Configuration files bundled with each dataset specify the parameter settings used in the experiments.

All datasets generated in this study are publicly available on the Open Science Framework (OSF) at \url{https://osf.io/3abkw} with DOI \url{https://doi.org/10.17605/OSF.IO/3ABKW}. The repository contains 84 simulation datasets, each corresponding to a specific configuration of model parameters.
For each dataset, we provide the simulation scripts and configuration files used to generate the results, enabling users to reproduce the datasets or modify parameters to create alternative simulation scenarios. In addition to the processed data, visualization outputs associated with each dataset are included to facilitate rapid inspection and analysis of simulation results.

\section{Conclusion}
\label{sec:conclusion}
We introduced an agent-based simulation framework that jointly models human mobility, social interactions, infectious disease transmission, and pathogen shedding to generate synthetic datasets for wastewater-based epidemiology. Because every shedding event is referenced in space and time, the resulting data make explicit how mobility redistributes pathogen shedding across a sewershed rather than fixing it at residences, capturing both individual-level dynamics and the aggregated environmental signals they produce. This enables controlled evaluation of outbreak detection, transmission analysis, and surveillance strategies under diverse behavioral and epidemiological scenarios, including the dynamic-catchment conditions that are difficult to study with real data.

By releasing 84 scenario configurations together with reproducible generation tools , this work lowers barriers to method development and benchmarking in mobility analysis, epidemiology, and wastewater surveillance. The framework also supports extension to other regions using publicly available geographic data, facilitating comparative and scalable experimentation. Future work includes coupling the generated shedding events with a physical model of in-sewer transport, calibrating the simulation against empirical multi-scale wastewater measurements and real-world mobility data, incorporating additional behavioral and intervention models. Together, these steps would move the framework from a primarily generative tool toward an inferential, dynamic-catchment approach of WBE, strengthening fine-scale disease surveillance and public health planning.

\section{Acknowledgement}

We acknowledge that we used ChatGPT 5.5 and Claude Opus 4.8 for code development and also to improve the language, writing quality, and clarity of this paper. 
We did not use the model to generate scientific contributions, experimental results, or technical claims. We also disabled model training on our inputs so that the revised text would not be used for future model training.
%

This publication was made possible by the Insight Net cooperative agreement CDC-RFA-FT-23-0069 from the CDC’s Center for Forecasting and Outbreak Analytics. Its contents are solely the responsibility of the authors and do not necessarily represent the official views of the Centers for Disease Control and Prevention. This research was also supported by NSF Grant \#2109647, Data-Driven Modeling to Improve Understanding of Human Behavior, Mobility, and Disease Spread.

\bibliographystyle{ACM-Reference-Format}
\bibliography{main}



\end{document}